\documentclass[fleqn,usenatbib]{mnras}

\usepackage{natbib}
\bibpunct{(}{)}{;}{a}{}{,}
\usepackage{graphicx}
\usepackage{txfonts}

\def\xmm{{\em XMM-Newton}}

\def\chan{{\em Chandra}}

\def\suzaku{{\em Suzaku}}
\def\nustar{{\em NuSTAR}}

\def \src {3A\,1954+319}

\title[\xmm\ and \nustar\ observations of 3A\,1954+319]{Accretion of a clumped wind from a red supergiant donor onto  a magnetar is suggested by the analysis of the \xmm\ and \nustar\ observations of the X-ray binary \src}
\author[E. Bozzo et al.]{
E. Bozzo,$^{1}$\thanks{E-mail: enrico.bozzo@unige.ch}
C. Ferrigno,$^{1}$
L. Oskinova,$^{2,3}$
and L. Ducci,$^{1,4}$
\\
$^{1}$Department of Astronomy, University of Geneva, Chemin d'Ecogia 16, CH-1290 Versoix, Switzerland\\
$^{2}$Institut f\"ur Physik und Astronomie, Universit\"at Potsdam, Karl-Liebknecht-Strasse 24/25, 14476 Potsdam, Germany\\
$^{3}$Kazan Federal University, Kremlevskaya Str., 18, Kazan, Russia \\
$^{4}$Institut f\"ur Astronomie und Astrophysik, Kepler Center for Astro and Particle Physics, University of T\"ubingen, Sand 1, 72076 T\"ubingen, Germany
}

\date{}

\begin{document}
\label{firstpage}
\pagerange{\pageref{firstpage}--\pageref{lastpage}}
\maketitle

\begin{abstract}
\src\ has been classified for a long time as a symbiotic X-ray binary, hosting a slowly rotating neutron star and an aged M red giant. Recently, this classification has been revised thanks to the discovery that the donor star is an M supergiant. This makes \src\ a rare type of high mass X-ray binary consisting of a neutron star and a red supergiant donor.
In this paper, we analyse two archival and still unpublished \xmm\ and \nustar\ observations of the source. We perform a detailed hardness ratio-resolved spectral analysis
to search for spectral variability that could help investigating the structures of the inhomogeneous M supergiant wind from which the neutron star is accreting. We discuss our results in the context of wind-fed supergiant X-ray binaries and show that the newest findings on \src\ reinforce the hypothesis that the neutron star in this system is endowed with a magnetar-like magnetic field strength ($\gtrsim10^{14}$~G).
\end{abstract}


\section{Introduction}
\label{sec:intro}

\src\ has been classified until recently as a rare symbiotic X-ray binary (SyXB), i.e.\ a binary system hosting an accreting neutron star (NS) and a red giant companion \citep[see, e.g.,][and references therein; hereafter TE14]{enoto14}. This system is known to host a slowly rotating NS \citep[spin period of about 5 hours;][]{corbet08} which has displayed alternate episodes of spin-up and spin-down \citep{marcu11}. The optical companion was identified as a M4-5 III star at a distance of 1.7~kpc about 20~years after the discovery of the source in X-rays by the \emph{Uhuru} satellite \citep{forman78apj}, following the provision of the arcsec-level localization by a \chan\ observation carried out in 2006 \citep{masetti06}. The orbital period of the system was never firmly established, although \citet{mattana06} reported a lower limit to the period of $\sim$400~days.

Recently, the classification of the optical companion of \src,\ as well as the distance to the source, has been revised by \citet{hinkle20}. Using the data from the Gaia mission \citep{gaia18} in combination with both high resolution near-IR spectroscopy and high time resolution optical lightcurves, it was convincingly shown that the optical counterpart of \src,\  is
a red supergiant (RSG). Its estimated mass is 9~$M_{\rm \odot}$ at a distance of $\sim$3.3~kpc. A new lower limit of 3~years was provided for the system orbital period based on a series of radial velocities obtained in the optical domain and covering part of the orbit (2.7~years).  \citet{hinkle20} suggested that \src\ is one of the only few known supergiant X-ray binaries (SgXBs)  containing a RSG donor rather than a much more typical blue supergiant donor \citep[BSG, see, e.g., ][for a recent review]{nunez17}.
\begin{figure*}
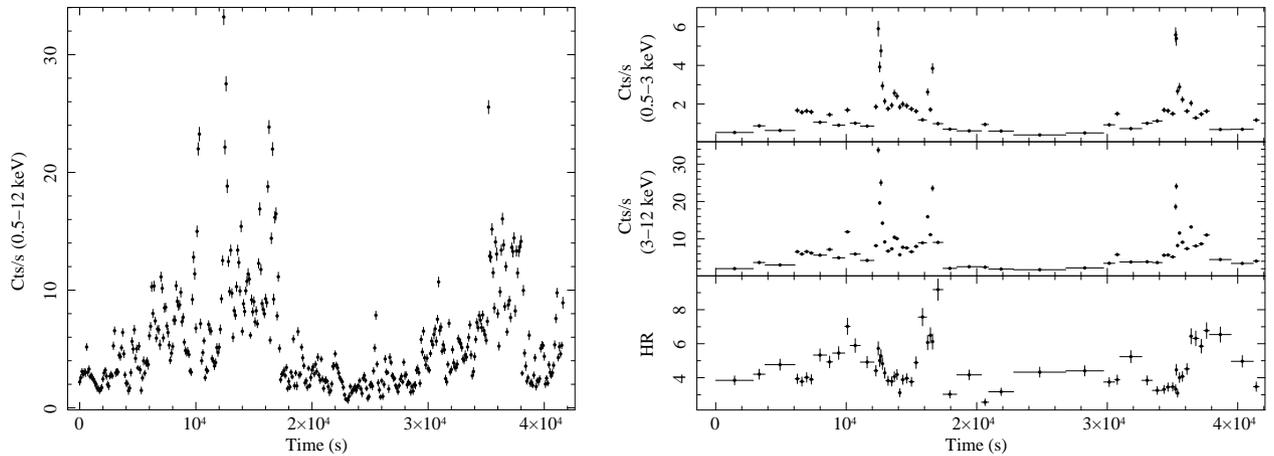

 \centering
 \includegraphics[width=5.9cm,angle=-90]{lc.ps}
 \includegraphics[width=5.9cm,angle=-90]{hr.ps}
  \caption{{\em Left}: 0.5-12~keV EPIC-pn light curve of \src\ with a time resolution of 100~s. {\em Right}: the energy-resolved EPIC-pn light curves of \src\ (top 0.5--3.5~keV and middle 3.5--10~keV  panels) and the correspondingly computed HR (bottom panel).}
  \label{fig:xmm_lc}
\end{figure*}

In this paper, we report on still unpublished \xmm\ and \nustar\ observation of the source. The \xmm\ observation is 43~ks long and was carried out in October 2011. The \nustar\ observation is 80~ks long and was carried out in June 2019. We perform a hardness ratio-resolved spectroscopic analysis of the \xmm\ and \nustar\ data, discussing the results in the context of wind-fed accreting supergiant high mass X-ray binaries. We also re-discuss at the light of the new classification of \src,\ the possibility that this system hosts a NS endowed with a magnetar-like magnetic field (TE14).

\section{Data analysis and results}

\subsection{\xmm\ data}
\label{sec:xmm}

\xmm\ observed \src\ on 2011-10-07 at 07:15 UT for a total exposure time of 43~ks (OBSID: 0675130201).
This corresponds to slightly more than two source pulse periods (see Sect.~\ref{sec:intro}).
All EPIC cameras were operated in timing mode. Data from the two grating instruments (RGS1 and RGS2)
were not usable given the large extinction in the direction of the source (see later in this section).
No flaring background intervals were recorded, and thus the entire exposure time available for the EPIC cameras could be used for the scientific analysis.
\begin{table}
\centering
\caption{\label{tab:xmm}Best-fit parameters obtained by using the EPIC-MOS1, MOS2, and pn data. The best fit model is obtained by using a power-law affected by a partial absorber. In the following, $N_\mathrm{H}$ is the absorption column density along the direction to the source (i.e. the Galactic absorption), $N_\mathrm{H, pc}$ is the absorption column density of the partial absorber (representing the absorption column density local to the source; {\sc pcfabs} in {\sc Xspec}), $f$ is the covering fraction of the partial absorber, $\Gamma$ is the power-law photon index, and F$_\mathrm{0.5-10~keV}$ is the measured power-law flux in the 2--10~keV energy range (not corrected for absorption).}
\begin{tabular}{lr@{}lll}

    \hline

    \hline
    \multicolumn{4}{c}{Average}\\
    \hline
    $N_\mathrm{H}$ & 2.57 &$\pm$0.05 & 10$^{22}$~atoms~cm$^{-2}$\\
    $N_\mathrm{H, p.c.}$ & 8.3 &$\pm$0.4 & 10$^{22}$~atoms~cm$^{-2}$\\
    $f$ & 0.618 &$\pm$0.011 & \\
    $\Gamma$ & 1.905 &$_{-0.023}^{+0.020}$ &  \\
    F$_\mathrm{2-10 keV}$ & 56.9 &$\pm$0.6 & $10^{-12}$cgs \\
    $\chi^2_\mathrm{red}$/d.o.f. &  1.1 &/309 & \\
    \hline
\end{tabular}
\end{table}

All observation data files (ODFs) were processed by using the \xmm\ Science Analysis System
(SAS 19.1.0) following standard procedures\footnote{http://www.cosmos.esa.int/web/xmm-newton/sas-threads}.
The regions used for the extraction of the source spectra and light curves were chosen
for all instruments to be centered on the best known position of \src\ (see Sect.~\ref{sec:intro}).
For the pn, we used all rows comprised between 25 and 49, while for the MOS2 the interval was 293-321.
For the MOS1, we excluded from the source product extraction region the rows adjacent to the dead pixel
row\footnote{https://www.cosmos.esa.int/web/xmm-newton/sas-watchout-mos1-timing} and thus used all rows in the ranges 302--316 and 322--338. We checked that including the dead pixel row while extracting the MOS1 spectra led to strong instrumental residuals at 1.7--2.3~keV, as well as below 1~keV. The residuals at the lower energies, not present in the MOS2 and pn data, appeared also in the products obtained from the split row extraction ranges mentioned above and thus these data were discarded for further analysis. The background spectra and light curve extraction regions for all cameras were chosen to lie in a portion of the instrument field of view free from the contamination of the source emission. We checked that reasonably different choices of the background extraction regions did not significantly affect the results. All lightcurves were corrected for the remaining instrumental effects by using the {\sc epiclccorr} task.

The background-corrected pn lightcurve of the source in the 0.5--12~keV energy range is reported in Fig.~\ref{fig:xmm_lc}. The source intensity during the \xmm\ observation (see Table~\ref{tab:xmm}), as well as the recorded source variability, are remarkably close to that of the 2011 \suzaku\ observation published by TE14. This was called by TE14 a ``quiescent'' emission state. As the \xmm\ data span a too short baseline to measure the spin period of the source (only two pulse cycles are observed), in the following we assume the reference value of 5.70$\pm$0.01~h measured by TE14.

In Fig.~\ref{fig:xmm_lc}, we also show the energy resolved pn light curves of the source and the corresponding hardness ratio (HR) between the bands 3.5--10~keV and 0.5--3.5~keV. This was computed with an adaptive rebinning \citep[see, e.g.,][]{bozzo13b}, achieving in each soft time bin a signal-to-noise ratio (S/N) of $\gtrsim$10. Prominent HR variations are recorded by the pn, suggesting the occurrence of  significant spectral changes during the course of the observation.

\begin{figure}
 \centering
 \includegraphics[width=6.4cm,angle=-90]{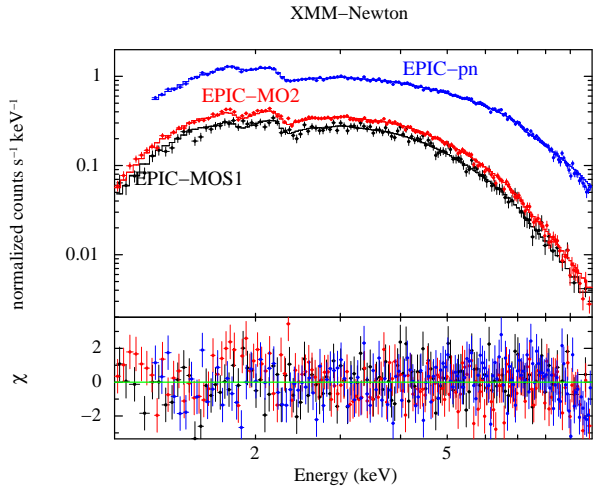}
  \caption{The three EPIC spectra obtained by using all time available during the observation of \src\ (OBSID: 0675130201). The best fit model is obtained by using an absorbed power-law plus a partial covering (see text for more details). The residuals from the fit are shown in the bottom panel.}
  \label{fig:xmm}
\end{figure}

We first extracted the average source spectra for the three EPIC cameras and fit them simultaneously. We used first a simple absorbed power-law model, adopting the {\em wilm} abundances \citep{wilms00} and {\em vern} cross sections \citep{vern96} to facilitate also the comparison with previous works in the literature on this source (see, e.g., TE14). This fit resulted in an unacceptable result, providing $\chi^2_{\rm red}$/d.o.f.=5.19/464. In agreement with TE14, we found a strong soft excess below 2 keV. An acceptable result could be obtained by including a partial absorber in the fit. This result is reported in Table~\ref{tab:xmm}. We favored the usage of a partial absorber ({\sc pcfabs} in {\sc Xspec}) over a blackbody thermal component as suggested by TE14 because the partial absorber is among the most commonly used soft component used to describe the X-ray emission from wind-fed SgXBs. This component is usually ascribed to the presence of a structured (``clumpy'') stellar wind that allows only part of the X-ray emission from the accreting NS to escape the bulk of the extinction caused by the dense and cold wind material. As the \xmm\ data cover only a limited energy band ($\lesssim$10~keV), we only verified a posteriori that a model comprising a powerlaw with a high energy cut-off ({\sc highecut*pow} in {\sc Xspec}) with values of $E_{\rm C}$ and $E_{\rm F}$ frozen to those measured from the \nustar\ data could not give an acceptable result unless the partial absorber is also included. However, in this case, there is virtually no improvement in the fit compared to that reported in Table~\ref{tab:xmm}. Normalization constants were included in all fits to take into account cross-calibrations between the pn, MOS1, and MOS2. Freezing to unity the normalization constant of the pn, those of the MOS1 and MOS2 turned out to be in all cases compatible with 0.8 (that is expected when all cameras are operated in timing mode, see  https://xmmweb.esac.esa.int/docs/documents/CAL-TN-0018.pdf). The average \xmm\ spectra, together with the best fit model and the residuals from the fit are shown in Fig.~\ref{fig:xmm}. We found no indication of the presence of an iron emission line at $\sim$6.4~keV in the EPIC spectra of \src.\ Adding a line at the same energy as that measured by \nustar\ (see Sect.~\ref{sec:nustar}), provides a 90\% c.l. upper limit on the line normalization of 5$\times$10$^{-6}$~photons~cm$^{-2}$~s$^{-1}$. The detection of iron lines in the X-ray spectra of \src\ has been shown already by TE14 to occur only during the brightest emission phases (i.e. during their 2012 \suzaku\ observation and not during the 2011 observation) and thus our findings here are fully in agreement with what has been reported previously for this source.

In this and in the following spectral analysis, we optimally grouped spectra following the prescriptions by
\citet{Kaastra}, minimized the modified Cash statistics (\texttt{Cstat}) in {\sc Xspec}, and computed parameter ranges using a Monte Carlo Markov chain with the Goodman-Weaver sampling algorithm (40 walkers, chain length 26\'000, and burning phase 6000). All uncertainties are reported at 68\% confidence interval using the corresponding quantiles in the posterior distribution. Priors are flat for power-law index and covering fraction, logarithmic for flux and absorption limits. We have checked the posteriors using corner plots to verify that priors were not influencing the parameter determination.

\begin{figure}
 \centering
 \includegraphics[width=\columnwidth,angle=0]{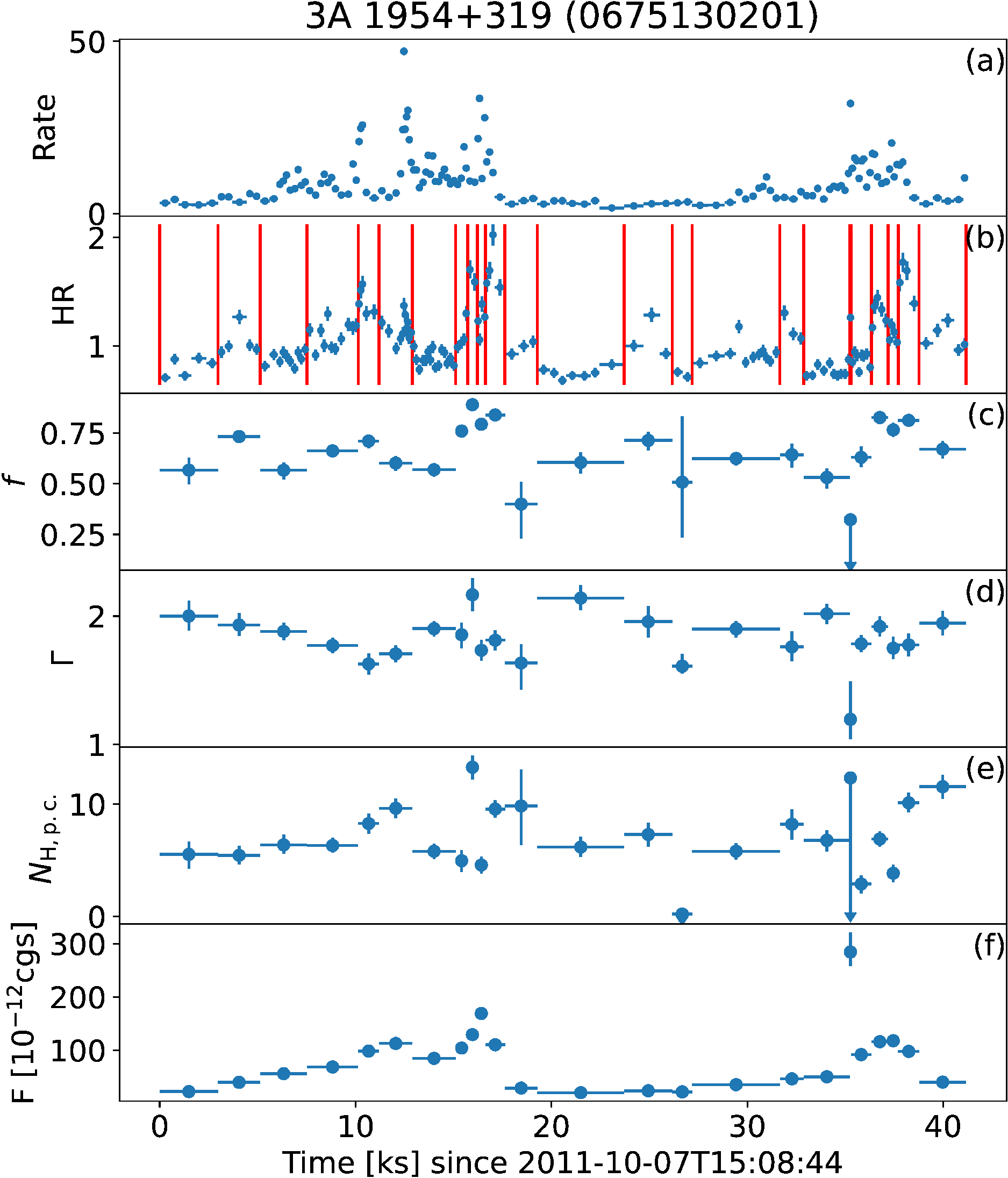}
  \caption{Results of the HR-resolved spectral analysis performed on the EPIC-pn data of \src.\ The different panels show the variations of the relevant spectral parameters (introduced in Table~\ref{tab:xmm}). In particular, panel (a) shows the source count-rate, panel (b) the recorded HR, panel (c) the covering fraction of the partial covering component, panel (d) the powerlaw photon index, panel (e) the absorption column density of the partial covering component, and panel (f) the source flux. The vertical red lines in panel (b) mark the different time intervals identified by the Bayesian block technique for the most relevant HR variations and used for the spectral extraction.}
  \label{fig:xmm_hr}
\end{figure}

In order to investigate the origin of the possible spectral variations suggested by the changes of the source HR in Fig.~\ref{fig:xmm_lc}, we performed an HR-resolved spectral analysis in analogy to our previous works on wind-fed SgXBs \citep[see, e.g.,][]{bozzo17,ferrigno20}. The HR-resolved time intervals used for the spectral extractions were determined by using the EPIC-pn data as reference and running the Bayesian Blocks analysis that we introduced and extensively illustrated in our previous paper \citep{ferrigno20}. These intervals are indicated in Fig.~\ref{fig:xmm_hr} through vertical red lines.
The MOS1, MOS2, and pn spectra of each HR-resolved  time interval were fit together with the same model adopted above for the time-average spectra. We fixed in all fits the value of the {\sc Tbabs} component to that determined from the average spectra (this is interpreted as the Galactic absorption and thus not supposed to change on the timescale covered by the \xmm\ observation). The results of this analysis are summarized in Fig.~\ref{fig:xmm_hr}. We verified a posteriori that fixing the value of $N_{\rm H}$ in all HR-resolved spectral fits did not change the qualitative and quantitative outcomes of this analysis.

\begin{figure}
 \centering
 \includegraphics[width=\columnwidth,angle=0]{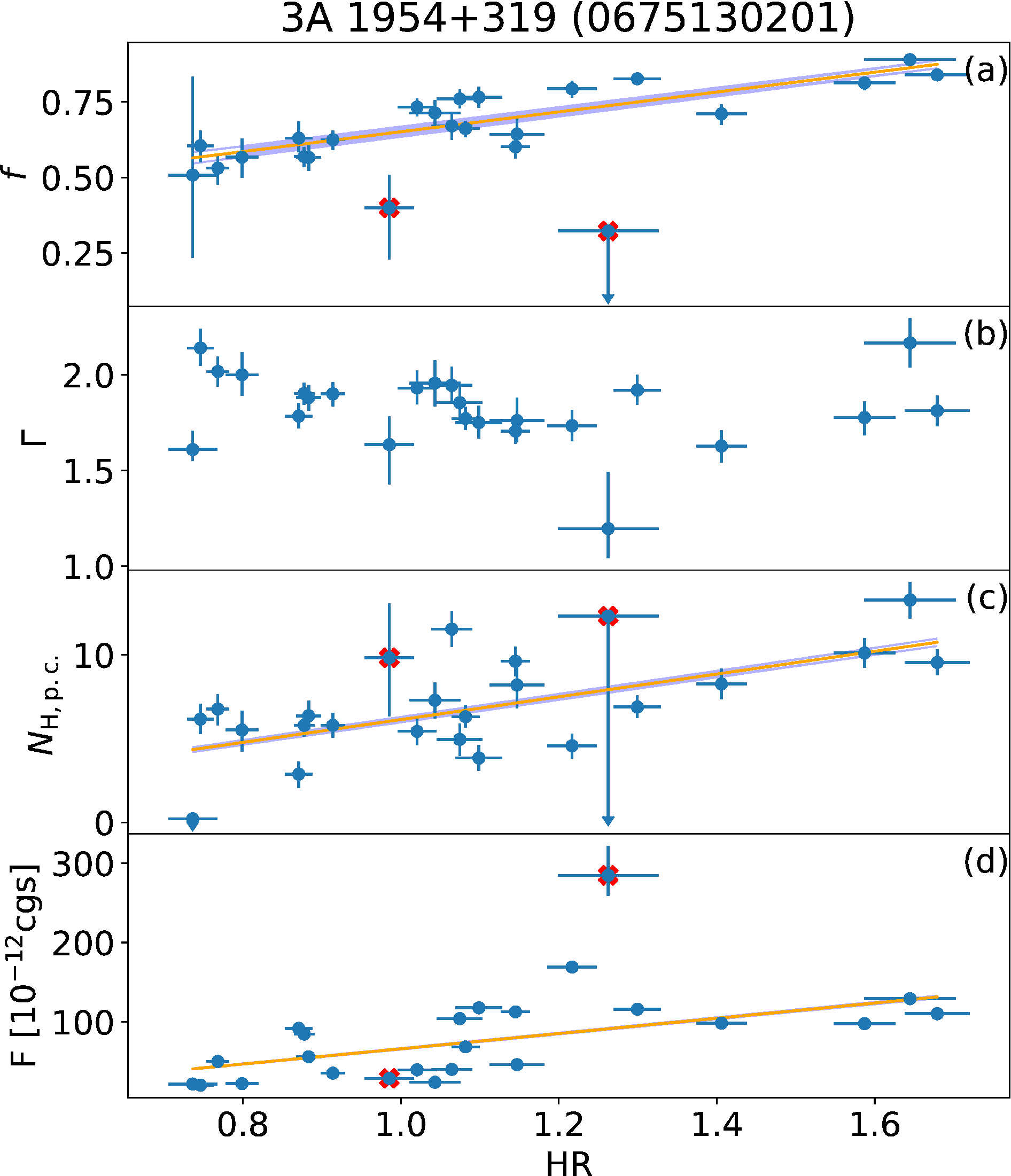}
  \caption{Plot of the different spectral parameters reported in Fig.~\ref{fig:xmm_hr} as a function of the source HR rather than time. In particular, panel (a) shows the covering fraction of the partial covering component, panel (b) the powerlaw photon index, panel (c) the absorption column density of the partial covering component, and panel (d) the source flux. In panels (a) and (c), corresponding to $N_{\rm H, pc}$ and $f$, we also show the determined linear correlations of these parameters as a function of the HR  and the envelope of the correlation curves at 1-$\sigma$ c.l. (shaded regions). We report also for reference the flux of the source as a function of HR. The two outliers in the correlations mentioned in the text are highlighted with red circles.}
  \label{fig:xmm_hr_spec}
\end{figure}

Figure~\ref{fig:xmm_hr} shows some interesting features. We observe that points corresponding to spikes in the HR, especially during the intervals of the brightest emission ($t=15-18$~ks and $t=35-40$~ks, where $t$ is the time measured from the beginning of the observation), are generally characterized by an increase of the covering fraction (by up to about 60\%) and the partial covering absorption column density (up to a factor of two). As we comment more extensively in Sect.~\ref{sec:discussion}, this behavior is reminiscent of what is usually observed in wind-fed SgXBs which show flares produced by inhomogeneities in the surrounding stellar wind \citep[see, e.g.,][and discussions therein]{bozzo17, ferrigno20}. These inhomogeneities, commonly addressed as ``clumps'' \citep[see, e.g.,][and reference therein]{nunez17}, produce an increase in the density around the compact object (thus resulting in an enhanced extinction of the X-rays) followed by an augmented mass accretion rate, in turns responsible for the brightening of the X-ray emission. Subsequently, the clump material gets photoionized by the NS emission and a decrease in the local absorption column density is observed. To show more clearly the identified correlations of the spectral parameters as a function of the HR, we produced the plots in Fig.~\ref{fig:xmm_hr_spec}. Here we can see that the covering fraction and the absorption column density of the partial covering component increases as a function of the HR. There seems to be only two ``outliers'' compared to the bulk of the points describing these correlations. The two outliers have been marked with red circles in Fig.~\ref{fig:xmm_hr_spec} and are characterized by a remarkably lower covering fraction compared to all other points (although the error bars of the different parameters associated to these points are also significantly larger than virtually all others). Furthermore, one of the two outliers is characterized by the highest achieved flux of the source across the entire \xmm\ observation. Later in this section, we comment separately to the source states corresponding to the two outliers. The powerlaw photon index remains relatively flat with values scattered equally around the mean value of $\Gamma=1.95$ for the entire range of the HR. We evaluated the statistical significance of the correlations of $N_{\rm H, pc}$ and $f$ as a function of the HR by using the same method we adopted in \citet{ferrigno20}. We throw a sample of 10\,000 realization assuming Gaussian distributions of uncertainties on both axes. We computed the linear regression parameters for each realization and found the 68\% percentiles of distributions of the null hypothesis probability that the slope is zero, based on the $r^2$ correlation coefficient\footnote{https://docs.scipy.org/doc/scipy/reference/ generated/scipy.stats.linregress.html}. In the case of the correlation of $f$ as a function of HR, the Pearson's coefficient 1-$\sigma$ interval in our bootstrapped realization is contained between 0.51 and 0.62, while the slope is 0.35$\pm$0.3 at 68\% c.l. In the case of the correlation of $N_{\rm H, pc}$ as a function of HR, the Pearson's coefficient 1-$\sigma$ interval in our bootstrapped realization is contained between 0.38 and 0.44, while the slope is 7.2$\pm$0.3 at 68\% c.l.

In the above calculations for the evaluation of the significance of the correlations, we ignored the two outliers mentioned in the previous paragraphs. The first outlier that is visible about 18~ks after the beginning of the observation in Fig.~\ref{fig:xmm_hr}, is characterized by a modest decrease in the local absorption compared to the preceding points on the left and a substantial drop of the partial covering fraction. As the outlier occurs during the sharp decrease in the source flux after some intense flares visible in the time interval $t=10-17~ks$, we argue that the properties of the first outlier can be interpreted in terms of a substantial reduction in the density of the material surrounding the NS from which the compact object was accreting. This seems reasonable also because during the following 15~ks of observation the source remains in a relatively low emission state. The second outlier that is visible about 35~ks after the beginning of the observation in Fig.~\ref{fig:xmm_hr}, is characterized by a dramatic flattening of the powerlaw photon index and only upper limits could be measured for the partial covering parameters. Simultaneously, the source reaches the highest recorded intrinsic flux during the \xmm\ observation. Although the statistics available during the time interval corresponding to this HR-resolved spectrum is relatively limited and the lack of further similar episodes observed by \xmm\ hampers our attempt to get to a firm conclusion, a reasonable interpretation of this event is that for a few hundreds of seconds the accreting compact object managed to photoionize most of the surrounding medium and swamped away all the inflowing material, getting directly exposed along the line of sight to the observer.

\subsection{\nustar\ data}
\label{sec:nustar}

\begin{figure*}
 \centering
\includegraphics[width=5.9cm,angle=-90]{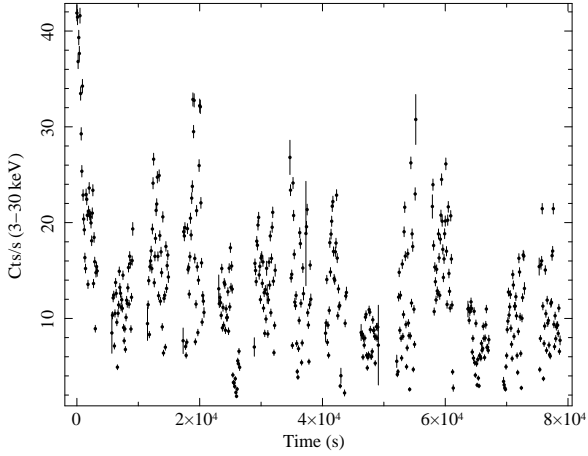}
\includegraphics[width=5.4cm,angle=-90]{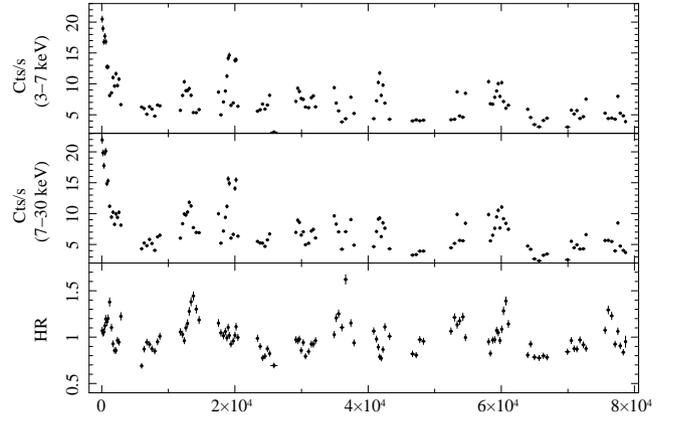}
  \caption{{\em Left}: 3-30~keV FPMA lightcurve of \src\ with a time resolution of 100~s. {\em Right}: the energy-resolved FPMA lightcurves of \src\ (top panel 3--7\,keV and middle panel 7--30\,keV) and the correspondingly computed HR (bottom panel; for the adaptive rebinning we used here in each soft time bin a S/N$\gtrsim$40). Only the results from the FPMA are shown for the sake of simplicity but we verified that the FPMB provided identical results within the statistical uncertainties.}
  \label{fig:nustar_lc}
\end{figure*}

\nustar\ observed \src\ on 2019-06-03 at 21:26 UT for a total spanned time of 80~ks (OBSID: 90501328002). This corresponds to slightly more than four source pulse periods (see Sect.~\ref{sec:intro}), although the effective exposure on the source for \nustar\ is roughly half of the time spanned by the observation due to the interruptions related to the orbit of the satellite (see Fig.~\ref{fig:nustar_lc}).
\begin{figure}
 \centering
 \includegraphics[width=6.4cm,angle=-90]{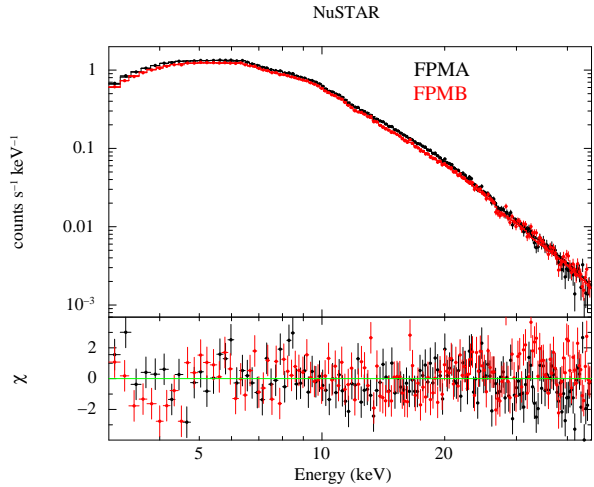}
  \caption{The FPMA and FPMB spectra of \src.\ The figure also shows the best fit model comprising an absorbed power-law with a high energy cut-off, plus a Gaussian iron line (see text for more details). The residuals from the fit are shown in the bottom panel.}
  \label{fig:nustar_spe}
\end{figure}

As for the \xmm\ data, also for \nustar\ we first extracted the energy-resolved lightcurves (3--7~keV and 7--30~keV) and computed the adaptively rebinned HR (see Fig.~\ref{fig:nustar_lc}). The intensity of the \nustar\ observation (see Table~\ref{tab:nustar_fit}), as well as the recorded source variability, are reminiscent of those visible in the 2012 \suzaku\ observation published by TE14 when the source was caught in a bright X-ray emission state. We then extracted the time-average FPMA and FPMB spectra of the source and attempted a combined fit using a simple absorbed power-law model. This fit gave an unacceptable result with a reduced $\chi^2\gtrsim$10. The introduction of a partial covering, as done in the case of the \xmm\ data, did not provide a significant improvement of the fit (leaving highly structured residuals especially above 10~~keV). We thus considered a model comprising a powerlaw with a high energy cutoff ({\sc highecut*pow} in {\sc Xspec}, see also TE14). This model provided a reasonably good fit to the data ($\chi^2/d.o.f.$=486.9/313), with structured residuals only evident around the energy of the iron line at $\sim$6.4~keV. This is a feature commonly observed in wind-fed systems. We thus included in the fit also a Gaussian line centered around the above energy and obtained the best fit parameters reported in Table~\ref{tab:nustar_fit}\footnote{The Goodman-Weaver chain was run with 60 walkers, burning phase of 26\,000, and length of 36\,000. The prior was logarithmic on the Iron line normalization and width, while it was linear for the centroid energy.}. A normalization constant was included in the fit to take into account cross-calibrations between the FPMA and FPMB but turned out to be fully compatible with unit. The FPMA and FPMB spectra, together with the best fit model and the residuals from the fit are shown in Fig.~\ref{fig:nustar_spe}.

\begin{table}
\centering
\caption{\label{tab:nustar_fit}Best-fit spectral parameters obtained from the time-averaged \nustar\ spectra of \src.\ The FPMA and FPMB spectra have been fit together. In the table below $E_\mathrm{Fe}$, $\sigma_\mathrm{Fe}$ $norm_\mathrm{Fe}$ are the centroid energy, width, and normalization of the Gaussian iron line. $E_\mathrm{C}$ and $E_\mathrm{F}$ are the cut-off and folding energies of the {\sc highecut} component in {\sc Xspec}. We also reported the flux calculated in the 2--10~keV energy range (not corrected for absorption) to ease the comparison with the \xmm\ data.}

\begin{tabular}{lr@{}ll}
    \hline
    \hline
$N_\mathrm{H}$ & 7.0 &$_{-0.4}^{+0.3}$ &  10$^{22}$~atoms~cm$^{-2}$\\
$E_\mathrm{Fe}$ & 6.19 &$_{-0.15}^{+0.10}$ & keV \\
$\sigma_\mathrm{Fe}$ & 0.4 &$_{-0.2}^{+0.3}$ &  \\
$\mathrm{norm}_\mathrm{Fe}$ & 2.8 &$_{-1.3}^{+2.3}$ & $10^{-4}$\,photons~cm$^{-2}$~s$^{-1}$ \\
$E_\mathrm{C}$ & 7.6 &$\pm$0.4 & keV\\
$E_\mathrm{F}$ & 23.7 &$\pm$0.7 & keV\\
$\Gamma$ & 1.599 &$_{-0.025}^{+0.020}$ &  \\
F$_\mathrm{2-10 keV}$ & 404 &$_{-8}^{+6}$ &  $10^{-12}$cgs \\
$\chi^2_\mathrm{red}$/d.o.f. &  1.5 &/442 & \\
    \hline
\end{tabular}

\end{table}

\begin{figure}
 \centering
 \includegraphics[width=\columnwidth,angle=0]{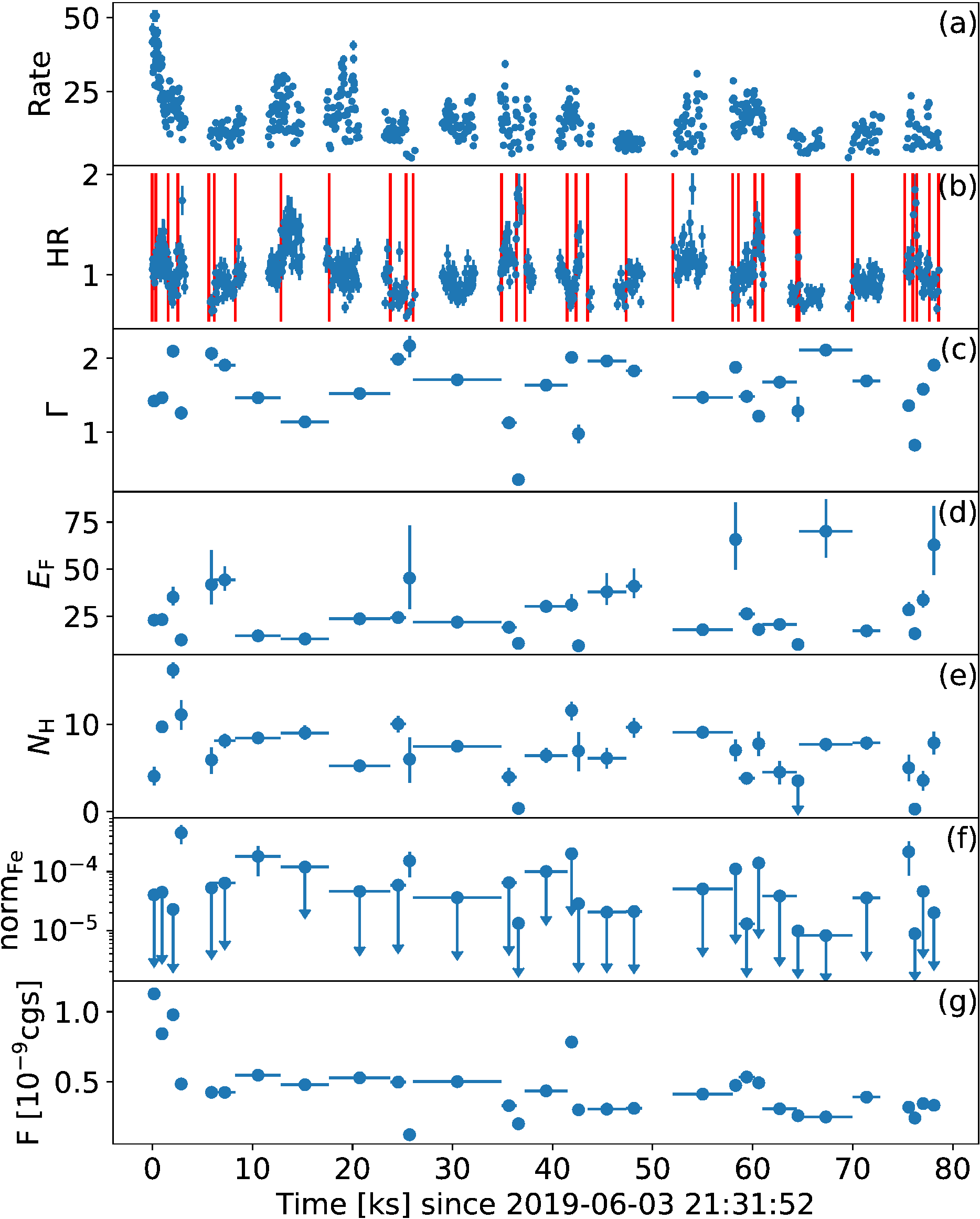}
  \caption{Results of the HR-resolved spectral analysis performed on the \nustar\ data of \src\ (using FPMA data as reference). The different panels show the variations of relevant spectral parameters (introduced in Tables~\ref{tab:xmm} and \ref{tab:nustar_fit}). In particular, panel (a) shows the source count-rate, panel (b) the recorded HR, panel (c) the powerlaw photon index, panel (d) the folding energy of the {\sc highecut} component, panel (e) the {\sc Tbabs} absorption column density, panel (f) the normalization of the iron line, and panel (g) the source flux. The vertical red lines in panel (b) mark the different time intervals identified by the Bayesian block technique for  the most relevant HR variations and used for the spectral extraction. The flux in panel (g) has been extrapolated from the fits in {\sc Xspec} in the 2--10~keV range to ease the comparison with the \xmm\ results (see Fig.~\ref{fig:xmm_hr_spec}).}
  \label{fig:nustar_hr}
\end{figure}

In order to investigate the origin of the possible spectral variations suggested by the changes of the source HR in Fig.~\ref{fig:nustar_lc}, we performed the same HR-resolved spectral analysis described before for the \xmm\ data (see Sect.~\ref{sec:xmm}). The results of the HR-resolved spectral analysis of the \nustar\ data is summarized in Fig.~\ref{fig:nustar_hr}. For each HR-resolved spectrum, the plot shows the changes in all relevant spectral parameters (in each HR-resolved time interval the FPMA and FPMB spectra were fit together to reduce the statistical uncertainties on the spectral parameters and the introduced normalization constants between the two instruments turned out to be compatible with unity in all cases). We fixed for all fits the values of the Iron line energy and width, as well as of the cut-off energy ($E_{\rm cut}$ in Table~\ref{tab:nustar_fit}) as these were largely constrained for the HR-resolved spectra and we verified {\it a posteriori} that fixing this parameter did not change the qualitative and quantitative  changes in all other parameters shown in Fig.~\ref{fig:nustar_hr}.

\begin{figure}
 \centering
 \includegraphics[width=\columnwidth,angle=0]{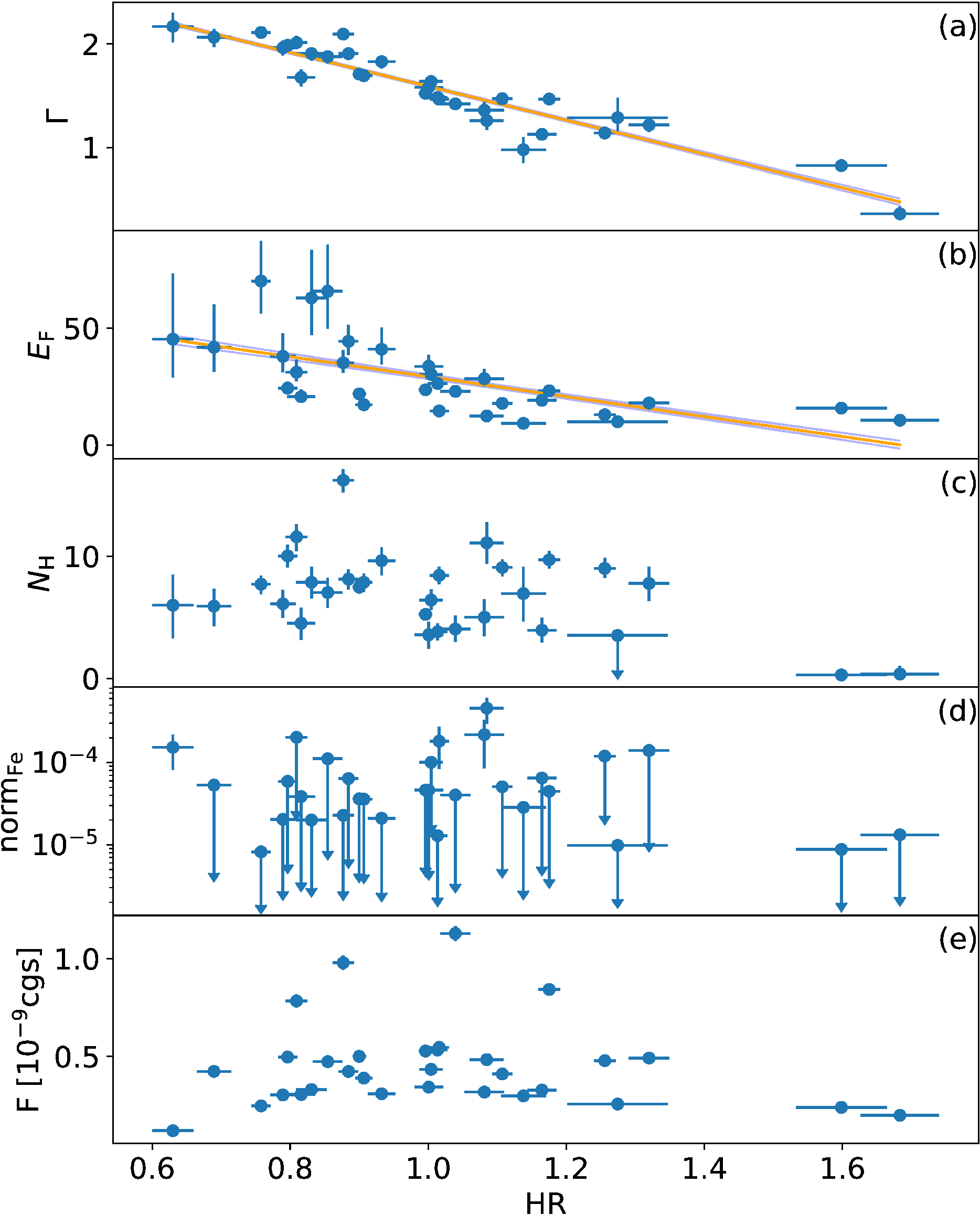}
  \caption{Plot of the different spectral parameters reported in Fig.~\ref{fig:nustar_hr} as a function of the source HR rather than time. In particular,  panel (a) shows the powerlaw photon index, panel (b) the folding energy of the {\sc highecut} component, panel (c) the {\sc Tbabs} absorption column density, panel (d) the normalization of the iron line, and panel (g) the source flux. In panels (a) and (b), corresponding to $\Gamma$ and $E_{\rm F}$, we also show the determined linear correlation and the envelope of the correlation curves at 1-$\sigma$ c.l. (shaded regions).}
  \label{fig:nustar_hr_spec}
\end{figure}

The results reported in Fig.~\ref{fig:nustar_hr} show some analogies with those obtained from \xmm\ (see Fig.~\ref{fig:xmm_hr}). At the beginning of the observation, when the source recovers from a bright flare, we notice a progressive increase in the local absorption column density that it is likely to result from the recombination of the photoionized material around the NS. The absorption column density remains then stable around roughly 8$\times$10$^{22}$~cm$^{-2}$. This might be related to the fact that the source was on average much brighter (a factor of $\gtrsim$100) during the \nustar\ observation compared to the \xmm\ one, providing a stable photoionization of the surrounding medium. The variations of the continuum emission are remarkable, but not clearly correlated with the source luminosity. We find, based on  Fig.~\ref{fig:nustar_hr_spec} that the hardness ratio is driven by both $\Gamma$ and $E_{\rm F}$. In the case of the correlation of $\Gamma$ as a function of HR, the Pearson's coefficient 1-$\sigma$ interval in our bootstrapped realization is contained between 0.30 and 0.38, while the slope is -1.63$\pm$0.04 at 68\% c.l. In the case of the correlation of $E_{\rm F}$ as a function of HR, the Pearson's coefficient 1-$\sigma$ interval in our bootstrapped realization is contained between 0.31 and 0.38, while the slope is -42.6$\pm$3.4 at 68\% c.l. No significant correlations are found for the absorption column density and for the normalization of the iron line (although, as shown in panel (d) of Fig.~\ref{fig:nustar_hr}, only for a few spectra the iron line is significantly detected, while in all other cases only an upper limit is determined).

\section{Discussion}
\label{sec:discussion}

The X-ray binary \src\ was recently reclassified from being a SyXB into a SgXB thanks to the discovery that the donor star in this system is an M-type supergiant. SgXB with RSG donors are rare: so far only two other RSG X-ray binaries are known in our Galaxy \citep[Scutum\,X-1
and CXO\,174528.79-290942.8; see][]{kaplan07,gottlieb20}. A handful of RSG X-ray binaries  identified in nearby galaxies are ultra-luminous X-ray sources (ULX) \citep[NGC 300 ULX-1, RX J004722.4-252051 in NGC 253, ULX J022721+333500 in NGC 925, and ULX J120922+295559 in NGC 4136,][]{heida15, heida16, heida19}. A few more possible candidates were reported by \citet{lau19}.

Due to the  proximity of \src, its X-ray flux is highest among the known system with RSG donors (at the best of our knowledge).
As such, given the larger number of X-ray photons detected per unit time, this system is particularly suitable to investigate the
the temporal and spectral variability of X-rays produced by the NS embedded in the  wind and accreting from it and thus
probe the RSG wind structure. Similar techniques are often used to study
the winds of BSG donors, especially to investigate the properties of clumps in winds of these hot stars \citep[][and references therein]{nunez17}.

The structure of dusty RSG winds is still poorly known, however the presence of clumps and structures is suggested by both the observational evidences of dense agglomerates rich in sub-structures \citep{kaminski19} and by the parallels with known properties of AGB stars \citep[somewhat less massive than RSG stars; see, e.g.,][]{agundez10, Decin2021}.
In order to conduct, for the first time, an investigation of the RSG wind structure using the wind accreting NS, we exploited two yet unpublished \xmm\ and \nustar\ observations of \src.  We looked for spectral  variability that could point towards the presence of massive and dense stellar wind structures around the compact object. The method we employ is based on the construction of adaptively rebinned HR plots and a Bayesian automatized way to select time intervals with significant variations of the HR to enhance the chance of finding the underlying variations of spectral parameters.

We aim at investigating changes in the continuum emission and the absorption column density in the vicinity of the X-ray source. Such changes suggest the presence of dense wind structures which cause X-ray flares when accreted and/or episodes of X-ray dimming when passing in front of the NS; this method is verified by our previous works on structures in the winds BSG donors \citep[see, e.g.,][]{bozzo11,bozzo16,bozzo17,pragati19,ferrigno20}. The results presented in this paper show
that the spectral variability observed in \src\ resembles that of other SgXB with BSG donors and occurs on similar timescales despite the fact that RSGs have winds that are on average a factor of $\sim$100 slower than BSGs. This is likely associated with the fact that the X-ray variability and the associated spectral changes emerge from the region close to the compact object that is dominating the dynamics of the accretion flow.

Our findings provide a new evidence of the RSG wind structuring and inhomogeneity. In particular, we show that structured flares from the source are preceded by raises of the local absorption column density which drops mainly close to the peak of the flares. This is expected if the flares are caused by dense structures approaching the compact object before being accreted and photoionized by the enhanced X-ray emission released during the accretion process \citep[see the detailed discussion in][]{bozzo11}.

As expected, these observational phenomena are more evident from the \xmm\ rather than the \nustar\ data. The first reason is that the source was in a much fainter X-ray luminosity state during the \xmm\ observation and thus the size of the ionized material around the object is smaller. This causes the flares to have a larger impact on the absorption column density variations that can thus be more easier revealed by X-ray observations. A second reason is that the higher energy threshold of the \nustar\ data does not allow us to reveal soft spectral components that are particularly critical to evaluate changes in the absorption column density local to the source (e.g., the partial covering component required by the \xmm\ data). The \nustar\ data revealed, however, interesting changes in the broad-band continuum of the source on short time scales. These changes are most likely associated to physical processes occurring close to the NS accretion column where the bulk of the X-ray radiation is originating. Unfortunately, the statistics  of the \nustar\ observation  are by far not sufficient for using our sophisticated spectral models that could be used to unveil changes in the accretion geometry \citep[see, e.g.,][and references therein]{ferrigno09,farinelli16}.

We remark here that the relatively short time-span of the \xmm\ and \nustar\ observations compared to the spin period of the NS in \src\ did not allow us neither to obtain new measurements of the NS spin period nor to perform a more detailed study of the spectral variations at different spin period phases. This latest analysis would ideally require the availability of many spin period cycles to wash-out occasional variations due, e.g., to the accretion environment, and was presented previously by TE14 using the longer time-span of the \suzaku\ observations. These \suzaku\ data cannot be used, on the other hand, to study the spectral variations that we looked for in the current paper. The instruments on-board \suzaku\ have a lower effective area than that available on the EPIC cameras and the FPMs (a factor of $\sim$5 for the \suzaku\,/XIS) and thus are not best suited for spectral analysis of moderately bright sources on time scales comparable to those used here (few hundreds to thousands of seconds, at the most). Future longer \xmm\ and \nustar\ observations of \src\ will be hopefully able to confirm and extend our findings, providing further insights on the presence of clumps in the winds of M supergiants and allow at the same time refined studies of the spectral variations at different spin period phases.

Beside opening a new window for studies of RSG stellar winds,
the renewed classification of the optical counterpart in \src\ as an RSG strengthens the possibility that this X-ray
binary might host a NS with a super-strong magnetic field \citep[in the magnetar regime; see, e.g.,][for a relevant review]{mereghetti16}.

As pointed out by \citet{hinkle20}, the newly determined source distance with Gaia implies that \src\ is twice as intrinsically luminous as previously thought. Furthermore, the velocity of stellar wind in the
RSG donor is roughly a factor of ten lower than the velocity previously assumed before for a red giant companion (about 10-30~km~s$^{-1}$ instead of a few hundreds of km~s$^{-1}$).
Thirdly, the orbital period of the system is now constrained to be at least of 3~years. Below, we follow the arguments presented by TE14 but update them with the new source distance, the corrected wind properties of the RSG donor, and the new lower limit on the system orbital period.

\begin{figure}
    \centering
    \includegraphics[width=\columnwidth,angle=0]{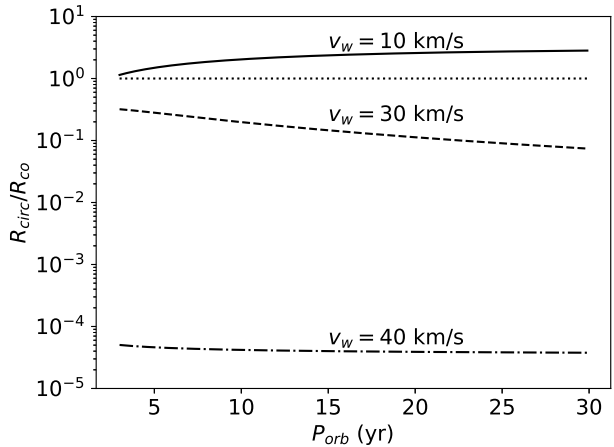}
    \caption{Plot of the circularization radius, $R_{\rm circ}$, in units of the corotation radius, $R_{\rm co}$, as a function of the orbital period of \src.\ The horizontal dotted line represents $R_{\rm co}$, while the solid, dashed and dot-dashed lines indicate $R_{\rm circ}/R_{\rm co}$ for different reasonable assumptions of the M supergiant companion hosted in \src\ (see text for more details).}
\label{fig:circ}
\end{figure}

As discussed by TE14, it is reasonable to assume that the NS in \src\ is close to the so-called spin-equilibrium and thus Eq.~5 in TE14 returns the highest possible NS magnetic field, $B$, if one assumes that the compact object in \src\ is accreting through an accretion disk. The value of $B\sim10^{16}~G$ obtained already by TE14 would not change much under the assumption of the updated system properties, as only the mass accretion rate enters the equation and it is increased by a factor of $\sim$2 (as this is directly proportional to the source intrinsic luminosity). A magnetic field as high as $B\sim10^{16}~G$ would be challenging also for a magnetar, but as noted by TE14, the hypothesis of \src\ being a disk accretor is poorly justified as for long orbital periods and slow wind velocities the formation of an accretion disk is unlikely. We formally verified here the above conclusion by TE14 checking if the angular momentum of the matter captured by the gravitational field of the NS in \src\ is sufficiently large to form an accretion disc. We used the concept of the circularization radius that for a wind-fed system is given by \citep{frank92apia}:
\begin{equation} \label{eq. rcirc}
  R_{\rm circ} \approx G^3 M_{\rm NS}^3 (v_{\rm orb}/a)^2 v_{\rm rel}^{-8}
\end{equation}
where $G$ is the gravitational constant, $M_{\rm NS}$ is the mass of the NS (assumed 1.4$M_\odot$), $v_{\rm orb}$ is the orbital velocity,
$a$ is the separation between the two stars, and $v_{\rm rel}$ is the relative velocity between the NS and the stellar wind from the supergiant. We assumed a circular orbit and we calculated $R_{\rm circ}$ as function of the orbital period, which is unknown but larger than 3 years, for three different values of the radial wind velocity in the correct range for an M supergiant. The results (see Fig.~\ref{fig:circ}) show that if we assume that the system is near the equilibrium, then the circularization radius is larger than the corotation radius only for very slow winds, enabling the formation of an accretion disk for a limited parameter space. We thus also conclude that the formation of an accretion disc in \src\ is unlikely.

A more reasonable and widely agreed scenario, as discussed earlier in this section and in several literature works on the source, is that \src\ is a wind accretor. In this case, the long equilibrium spin period of the source can be explained by assuming the presence of an extended subsonic (or ``settling'') accretion regime, where the NS is accreting from a surrounding shell of hot material.
The equilibrium spin period achievable in the settling regime was discussed extensively by \citet{shakura12} and TE14 showed all dependencies of this period in their Eq.~6\footnote{Note that the same approach and the same equation was discussed also by \citet{marcu11}.}. TE14 showed that the adoption of the settling accretion regime, given the relatively low X-ray luminosity of the source ($\lesssim4\times10^{36}~erg~s^{-1}$), would require a rather typical magnetic field for a young pulsar in a sgXB ($\sim$10$^{12}$-10$^{13}$~G) to achieve a period of roughly 5.7~h, thus excluding the magnetar hypothesis.

At the newly estimated distance of 3.3~kpc, the largest recorded X-ray luminosity by either \nustar\ or \suzaku\ would still be significantly lower than the limit above for the application of the settling accretion regime, thus validating the treatment adopted by TE14. However, the newly estimated lower limit on the source orbital period and the one order of magnitude reduction in the wind velocity of an M supergiant compared to a red giant would now make it impossible to avoid the magnetar case even in the assumption of the wind accretion case. By using a mass accretion rate of $\dot{M}_{16}=0.8$, an orbital period of $P_{\rm orb}=1095~days$ (lower limit), a relatively high wind velocity of $v_{\rm w}=40~km/s$, and a spin period of $P_{\rm spin}=20520~s$, the obtained magnetic field would be this time about 2$\times$10$^{16}$~G \citep[see also][]{cn21}. As for the case of disk accretion, this magnetic field would be very difficult to achieve even in the most extreme magnetars\footnote{Note that increasing the system orbital period helps reducing the required NS magnetic field strength, and the adopted value of 3~years has been indicated by \citet{hinkle20} as a lower limit. However, the dependence from the orbital period in Eq.~6 of TE14 is weak and thus the orbital period should be increased by at least 1-2 orders of magnitudes to recover a more reasonable magnetar-like magnetic field strength for \src.\ } but it is correct to remark here that such value should not be taken literally. First of all, the settling accretion regime and the corresponding spin-up/spin-down torques onto the NS have never been applied to magnetar-like fields and the limitations of the theoretical calculations should be revised carefully. As an example, the current treatment of the settling accretion regime \citep{shakura12} does not include the effects of magnetic and centrifugal gatings that have been discussed by \citet{bozzo08} and that become highly relevant when dealing with highly magnetized compact objects. In the present case, a magnetic field above 10$^{16}$~G, for example, would cause \src\ to be permanently in a strong propeller regime as the NS magnetospheric radius would be roughly 4 times larger than the corotation radius (see Eq.~8 and 9 in TE14). Therefore, \src\ could hardly be accreting at all, in contrast with observations. The calculations reported previously by \citet{bozzo08} did not include an extensive treatment of the accretion/ejection torques acting on the NS when gating mechanisms are at work and thus it is not straightforward to compute a modified version of Eq.~6 in TE14 for the expected equilibrium spin period.

Our conclusion is that \src\ is proved as of today to likely host a very strongly magnetized NS but a more realistic estimate of the compact object magnetic field strength requires a self consistent development of its spin history through appropriate spin-up and spin-down torques that is beyond the scope of the current paper and will be reported elsewhere.

\section*{Acknowledgements}
We thank the anonymous referee for detailed comments that helped us improve the paper.

\section*{Data availability}
The data underlying this article are publicly available from the \xmm\ and \nustar\ archives. The full chain of the analysis of the \xmm\ data is available in the form of a python Notebook that can be run in a dockerized environment at \url{https://gitlab.astro.unige.ch/ferrigno/4u1954-xmm.git}, the \nustar\ analysis can be reproduced using the notebook at \url{https://gitlab.astro.unige.ch/ferrigno/4u1954-nustar.git}.

\bibliography{bib.bib}{}
\bibliographystyle{mnras}

\end{document}